\documentclass[twocolumn,showpacs,prl]{revtex4-1}
\usepackage{graphicx}
\usepackage{color}
\usepackage{amssymb}

\newcommand{\abs}[1]{\left| #1\right|}
\begin{document}
\title{Observation of a hadronic interference effect in annihilation processes}
\author{Eef van Beveren$^{1}$ and George Rupp$^{2}$}
\affiliation{
$^{1}$Centro de F\'{\i}sica Computacional,
Departamento de F\'{\i}sica, Universidade de Coimbra,
P-3004-516 Coimbra, Portugal\\
$^{2}$Centro de F\'{\i}sica das Interac\c{c}\~{o}es Fundamentais,
Instituto Superior T\'{e}cnico, Technical University of Lisbon,
P-1049-001 Lisboa, Portugal
}
\date{\today}

\begin{abstract}
We present evidence for small oscillations
we observe in $e^{+}e^{-}$ and $p\bar{p}$ annihilation data,
with a periodicity of 76$\pm$2 MeV, independent of the beam energy.
We discuss some possible scenarios to explain the phenomenon.
\end{abstract}


\maketitle

\begin{figure}[htbp]
\begin{center}
\begin{tabular}{c}
\includegraphics[height=120pt]{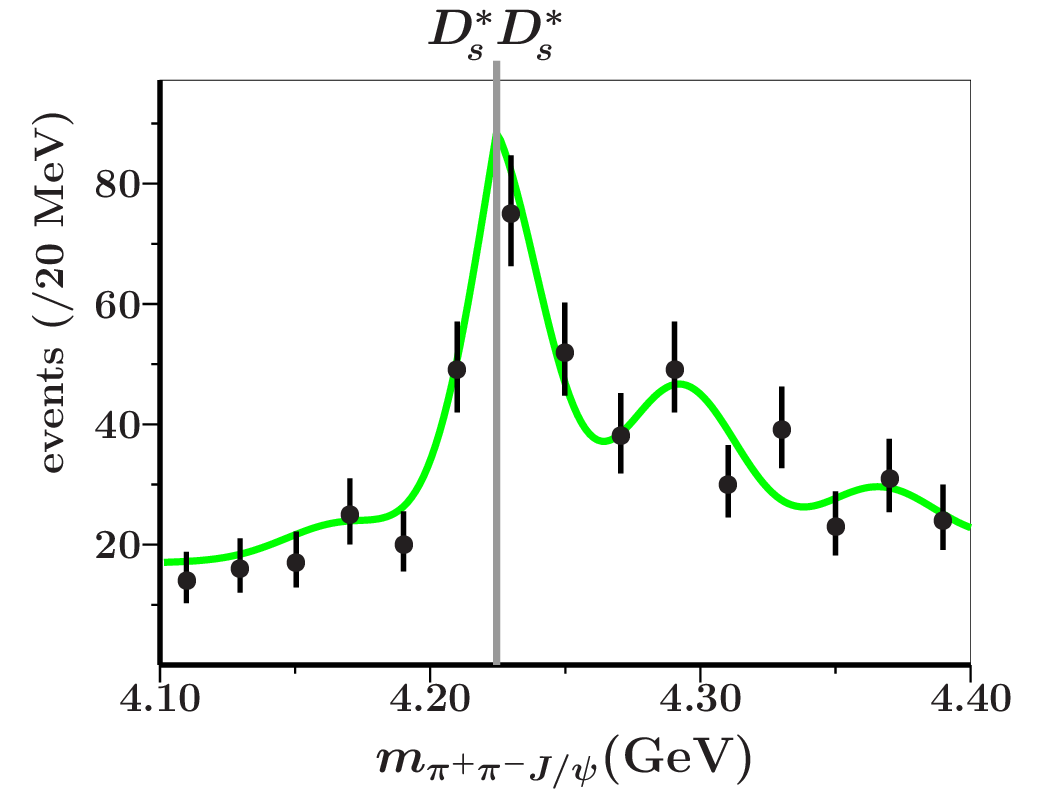}\\ [-10pt]
\end{tabular}
\end{center}
\caption{\small
Interference effect around the $D_{s}^{\ast}\bar{D}_{s}^{\ast}$
threshold in $e^{+}e^{-}\to J/\psi\pi^{+}\pi^{-}$ data
of the BABAR Collaboration \cite{ARXIV08081543},
with a periodicity of 74 MeV.
}
\label{X4260}
\end{figure}

In Ref.~\cite{PRD79p111501R},
we made notice of an apparent interference effect, which we observed
in the recent preliminary radiation data of the BABAR Collaboration
\cite{ARXIV08081543} (see Fig.~\ref{X4260}).
The effect, with a periodicity of about 74 MeV,
may be due to interference between the typical oscillation frequency
of the $c\bar{c}$ pair and that of the gluon cloud.
This interference effect is still awaiting confirmation by experiment,
which inevitably would require a binning smaller
than the 20 MeV of the actual data \cite{ARXIV08081543}.

In the meantime, we observe a very similar effect
in rather accurate data on $p\bar{p}\to J/\psi\pi^{+}\pi^{-}$
around the mass of the $X(3872)$ resonance,
\begin{figure}[b]
\begin{center}
\begin{tabular}{c}
\includegraphics[width=240pt]{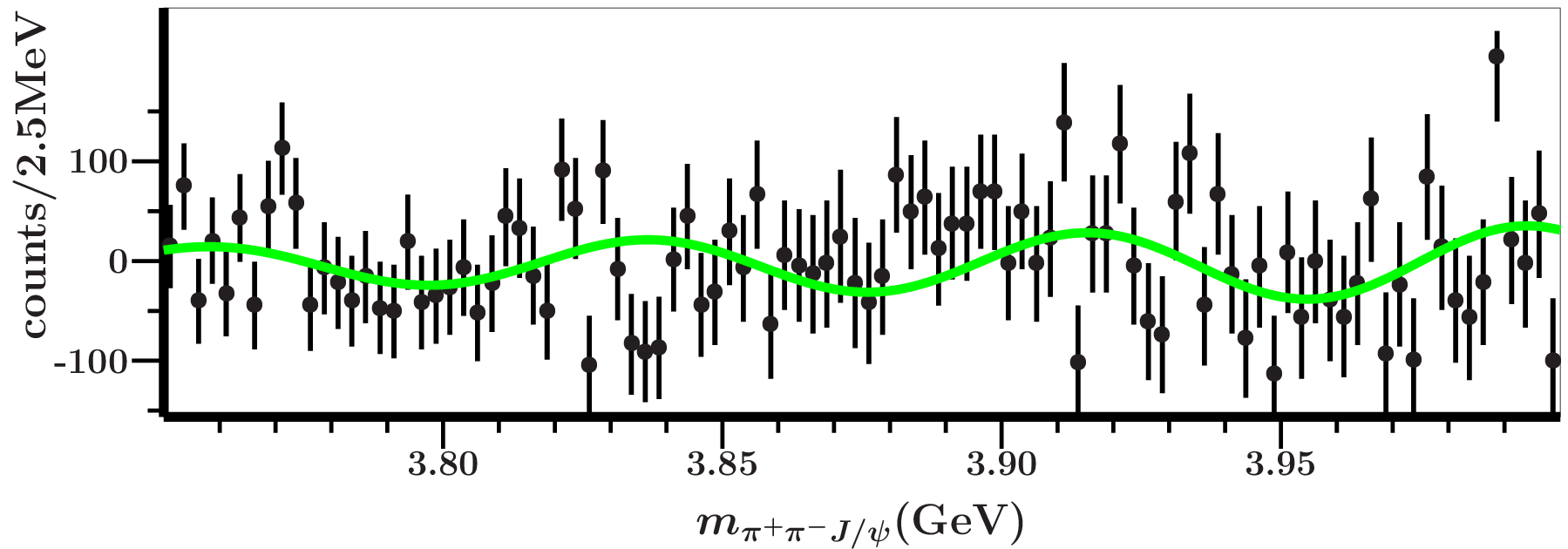}\\ [-10pt]
\end{tabular}
\end{center}
\caption{\small
Best fit
to the residual $p\bar{p}\to J/\psi\pi^{+}\pi^{-}$ data
of the CDF Collaboration \cite{PRL103p152001}, with a period of 79 MeV.
}
\label{cdf}
\end{figure}
obtained by the CDF Collaboration \cite{PRL103p152001}
(see Fig.~\ref{cdf}).
Here, we find that the period equals about 79 MeV.
However, we must allow for an uncertainty of roughly 4--5 MeV
in the periodicity observed in these data,
since CDF estimated their accuracy
on the spreading in invariant mass
by assuming a width of 1.3 MeV for the $X(3872)$ resonance,
whereas their signal width appears to be about 10 MeV.

\begin{figure}[htbp]
\begin{center}
\begin{tabular}{c}
\includegraphics[height=120pt]{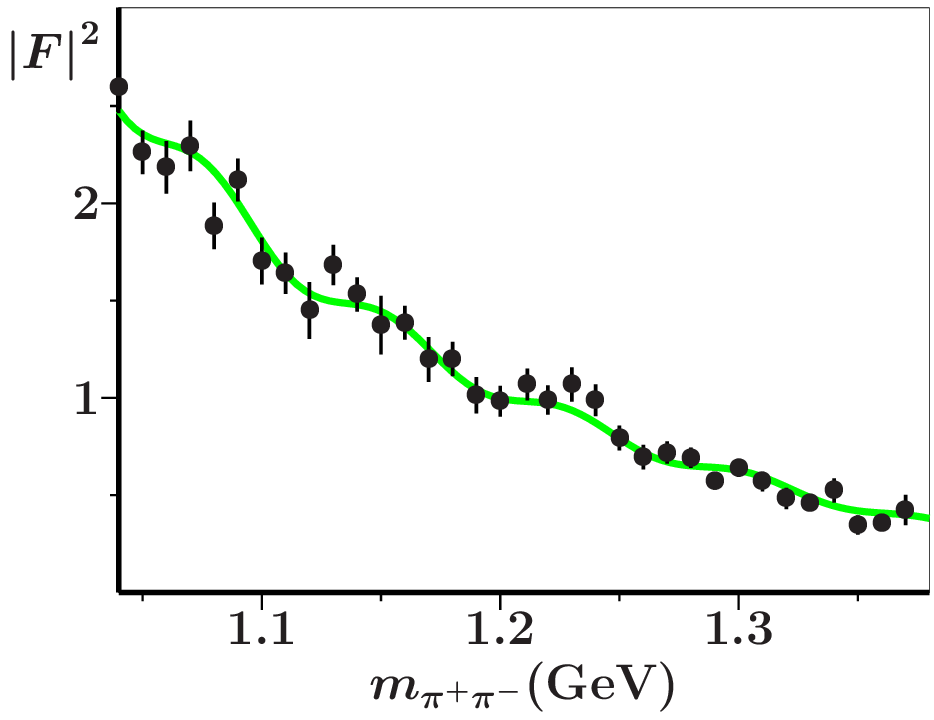}\\ [-10pt]
\end{tabular}
\end{center}
\caption{\small
Best fit
to the $e^{+}e^{-}\to\pi^{+}\pi^{-}$ data of the CMD-2 Collaboration
\cite{JETPL82p743}, with a period of 75 MeV.
}
\label{cmd2}
\end{figure}

Such an interference effect can also be observed in data
taken by the CMD-2 Collaboration for $e^{+}e^{-}\to\pi^{+}\pi^{-}$
\cite{JETPL82p743} (see Fig.~\ref{cmd2}),
with a periodicity of about 75 MeV and an estimated uncertainty of
some 2 MeV.

\begin{figure}[htbp]
\begin{center}
\begin{tabular}{c}
\includegraphics[height=120pt]{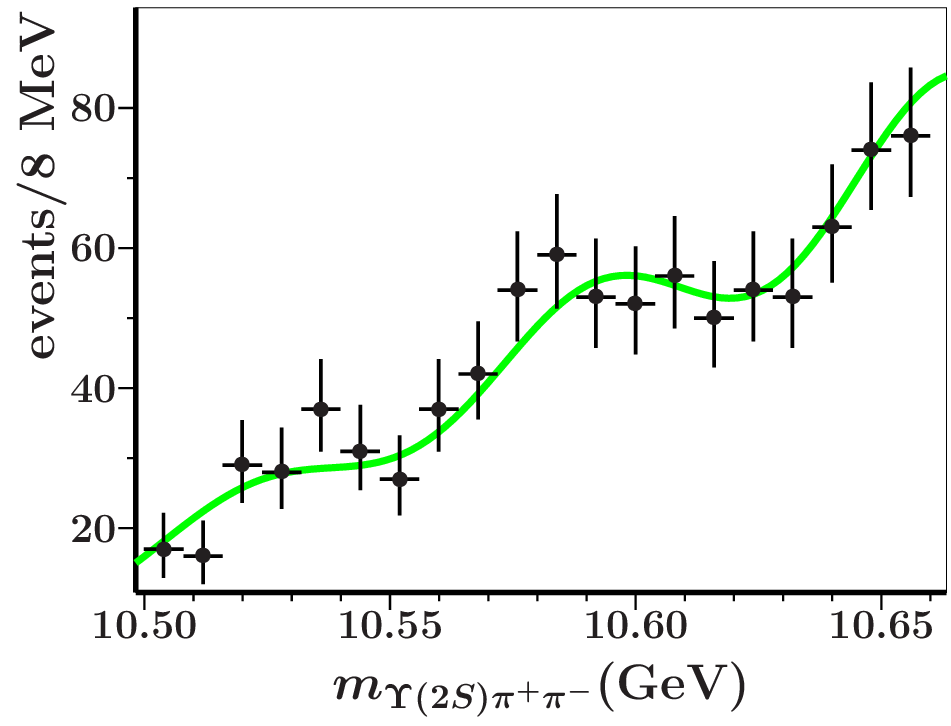}\\ [-10pt]
\end{tabular}
\end{center}
\caption{\small
Best fit to the $e^{+}e^{-}\to\Upsilon (2S)\pi^{+}\pi^{-}$ data
of the BABAR Collaboration \cite{PRD78p112002},
with a period of 73 MeV.}
\label{babar}
\end{figure}

Finally, data from the BABAR Collaboration
for $e^{+}e^{-}\to\Upsilon (2S)\pi^{+}\pi^{-}$
\cite{PRD78p112002}
show similar oscillations, with a periodicity of 73$\pm$3 MeV
(see Fig.~\ref{babar}).
However, note that the BABAR Collaboration
has a very different interpretation of the same data.
So it is questionable whether these data can be used in this context.
Nevertheless, as this is the only example we are aware of with sufficient
statistics for the observation of an oscillation in the $b\bar{b}$ sector,
we leave the interpretation as a question open to debate.

All fits have been done using a simple cosine,
with argument linearly proportional to the invariant mass
and amplitude proportional to the average magnitude of the signal.
An exception is the fit to the data of Fig.~\ref{cdf},
for which we have used the magnitude of the background signal
in the resonance region.

The first striking property of these inferred periodicities
is that it is quite constant, independent of the respective process,
viz.\ $e^{+}e^{-}\to J/\psi\pi^{+}\pi^{-}$,
$p\bar{p}\to J/\psi\pi^{+}\pi^{-}$,
$e^{+}e^{-}\to\pi^{+}\pi^{-}$,
$e^{+}e^{-}\to\Upsilon(2S)\pi^{+}\pi^{-}$,
as well as the flavors involved,
namely $n\bar{n}$ ($n=u$ or $d$),
$c\bar{c}$, and possibly also $b\bar{b}$.
From the observed values and the uncertainty estimates,
we feel safe to conclude that the periodicity has
a value of 76$\pm$2 MeV.

Observables that are independent of flavor
must somehow be related to gluons.
One such observable is the level spacing for quarkonium spectra
(see Ref.~\cite{AP324p1620} and references therein).
Here, we seem to have observed a second one.

Furthermore, the effect  seems to be quite  small,
but larger for $e^{+}e^{-}$ than for $p\bar{p}$ annihilation,
where it easily passes unnoticed
(see Fig.~\ref{cdf}).
In the latter case, the amplitude of the oscillation
is just one percent of the total signal,
whereas for $e^{+}e^{-}$ we find about 10 percent.
Anyhow, these numbers must yet be confirmed by further experiments.

Scenarios for an explanation of the phenomenon
can be split into two classes:
either some process which takes place before annihilation,
or interference of two different processes, after annihilation, that lead
to the same decay mode.
For the former class we have at present no serious candidate,
especially because the annihilation mechanisms
for $e^{+}e^{-}$ and $p\bar{p}$ are very different.
However, in the post-annihilation case, involving the formation of
$c\bar{c}$ in $e^{+}e^{-}\to J/\psi\pi^{+}\pi^{-}$
and $p\bar{p}\to J/\psi\pi^{+}\pi^{-}$,
$n\bar{n}$ in $e^{+}e^{-}\to\pi^{+}\pi^{-}$,
or $b\bar{b}$ in $e^{+}e^{-}\to\Upsilon(2S)\pi^{+}\pi^{-}$,
we will present here a possible scenario.

For the decay of charmonium into $J/\psi\pi^{+}\pi^{-}$,
one expects that the reaction
is dominated by the transition to a stable charmonium state,
$J/\psi$ or $\psi (2S)$,
through peripheral emission \cite{PRL105p102001}
of a $\sigma$- or $\rho$-like structure
which then decays into two pions.
But it is also possible that emission takes place from the interior
of the charmonium state, clsoe to the $c\bar{c}$ pair.
However, the latter reaction is much less probable,
as pair creation near the $c\bar{c}$ pair
dominantly leads to open-charm decay.
For the process $e^{+}e^{-}\to J/\psi\pi^{+}\pi^{-}$,
this seems a sufficient explanation for the existence
of two distinct reactions which might interfere.
On the other hand, in $p\bar{p}\to J/\psi\pi^{+}\pi^{-}$,
many other reactions may take place, for instance
annihilation of just one light $q\bar{q}$ pair and subsequent creation of
a $c\bar{c}$ pair,
with a rearrangement of the remaining light quarks and antiquarks.
Full annihilation of $p\bar{p}$ will probably only take place
for a very small fraction of the events.
In case our estimates of the amplitudes of the oscillation
are correct, we find that full $p\bar{p}$ annihilation
only occurs in 1 out of 10 events.

So far charmonium and possibly beautonium,
but for the process $e^{+}e^{-}\to\pi^{+}\pi^{-}$
we must assume that, after the creation of an initial $n\bar{n}$ pair,
the OZI-allowed reaction $n\bar{n}\to (n\bar{q})+(q\bar{n})$
($q$ also light) dominates.
Nevertheless, non-OZI reactions will also take place.
Consequently, also in this case we may expect
interference from the two different reactions.

Quantum interference of particles and resonances
was recently studied by Ya.~Azimov
\cite{JPG37p023001}.
In his paper he reminds:
``{\it Regretfully, the structure of both the rescattering interference
and different interference effects in decays is not yet clearly understood.
That is why fits to experimental data are
still very model-dependent in many cases.}''
However, in our fits for the present cases,
we understand that no model dependence
has slipped yet into our observations.

Oscillations have been reported by
S.~Pacetti \cite{FRWP013} in diffractive photoproduction data
obtained in the E687 experiment at Fermilab \cite{PLB514p240},
though with a periodicity of about 250 MeV in momentum transfer.
The author concludes
``{\it We find at least five interfering structures, but
to have a clear identification of this (sic) resonances, we need
much more precise data.}''
In Ref.~\cite{PLB397p305},
P.~Gauron, B.~Nicolescu, and O.~V.~Selyugin
demonstrate that the high-precision $dN/d\abs{t}$, in $p\bar{p}$ data
collected by the UA4/2 Collaboration
at the CERN S$\bar{p}p$S Collider at $\sqrt{s}=541$ GeV
\cite{PLB316p448},
shows oscillations at very small momentum transfers.
These oscillations seem to be periodic in $\sqrt{t}$,
with a periodicity of about 20 MeV.
Oscillations of the hadronic amplitude at small transferred momenta
are discussed by O.~V.~Selyugin in Ref.~\cite{Diffraction95p65},
while S.~Barshay and P.~Heiliger signal \cite{ZPC64p675}
signs for new physics from oscillating behaviour
in the amplitude of hadronic diffractive scattering data.
They emphasize:
``{\it possible signals coming directly from such
a new condition of matter, that may be present in current
experiments on inelastic processes.}''
Fourier analysis of oscillations in hadronic amplitudes
is performed by J.~Kontros and A.~Lengyel \cite{HADRONS96p186},
but the periodicity in $\sqrt{\abs{t}}$ of their oscillations
is two orders of magnitude larger
than the oscillations considered in Ref.~\cite{PLB397p305}.
Furthermore, in Ref.~\cite{PNL5p114}
Y.~A.~Troyan and collaborators observe oscillations
in $\pi^{+}\pi^{-}$
from the reaction $np\to np\pi^{+}\pi^{-}$ at $P_{n}=5.20$ Gev/c.
Unfortunately, the data are not well represented,
and so do not allow to extract the periodicity.

For the present observation of a constant periodicity in $\sqrt{s}$,
we keep on assuming that it occurs because of two distinct
processes, namely peripheral emission and
pair creation in the deep interior of a meson.
In the past, we have shown that the oscillations
of quarkonia are independent of flavor
and have a frequency $\omega =190$ MeV
(see Ref.~\cite{AP324p1620} and references therein).
Upon pair creation in the interior,
the signal most probably picks up this frequency.
Emission, which we assume to originate from the gluon cloud,
would initially keep the gluon frequency.
Therefore, if we assume that the periodicity of 76$\pm$2 MeV
observed in annihilation processes stems from interference
between the two signals, then we are led to frequencies,
for gluon oscillations, of either $190+2\times (76\pm 2)=342\pm 4$ MeV
or $190-2\times (76\pm 2)=38\pm 4$ MeV.
The former value would give rise to radial gluon excitations
with level spacings of about 684 MeV, which is in reasonable
agreement with the level spacings from the lattice
obtained by C.~J.~Morningstar \& M.~Peardon
in Ref.~\cite{PRD60p034509},
by Liu \& Chuan in Ref.~\cite{CPL18p187}, and
by E.~B.~Gregory and collaborators in Ref.~\cite{HEPLAT0510066}.
An extensive discussion on glueballs can be found in
Ref.~\cite{PREP454p1} by E.~Klempt and A.~Zaitsev.

At first sight, we are inclined to reject
the value of $38\pm 4$ MeV for the frequency
of gluon oscillations in mesonic configurations.
But from Anti-de-Sitter (AdS) confinement \cite{NCA80p401},
we learn that gluons and quarks all oscillate
with the same frequency,
which is given by the radius of the AdS system.
Hence, neither of the two solutions seems to be acceptable then,
because for quarkonia spectra
we obtain excellent results with $\omega =190$ MeV.
However, AdS confinement also indicates that
the gluon distribution is concentrated towards the surface,
much like in the bag model.

This opens up the possibility of surface oscillations,
with level spacings that are equal to the oscillation frequency,
and which have been solved for the bag model
by T.~A.~DeGrand and C.~Rebbi in Ref.~\cite{PRD17p2358}.
Actually, they found frequencies
for the lowest-order surface vibrations
very comparable to the value $342\pm 4$ MeV.
However, in Ref.~\cite{PRD27p2708},
P.~J.~Mulders and collaborators
obtained for light baryons values
in the range 0.38--0.54 GeV,
and for $b\bar{b}$ quarkonia values even higher than 0.6 GeV,
which all seem to be well beyond our result.
In Ref.~\cite{PRD31p2902}, H.~R.~Fiebig found for
the inertia of surface oscillations in light baryons
a mass parameter, which,
when related by the expressions given in Ref.~\cite{PRD27p2708}
to the oscillation frequency, gives $\omega =0.38$ GeV,
in reasonable agreement with our value for mesons above.
In Ref.~\cite{PRC71p065203}, H.~P.~Morsch and P.~Zupranski
found evidence for a breathing mode of the nucleon,
the ``scalar'' $P_{11}$ excitation at 1.4 GeV with a width of 0.2 GeV,
from high-energy proton-proton scattering.
Finally, in Ref.~\cite{NUCLEX0612015},
the CELSIUS-WASA Collaboration
find similar evidence from the large branching fraction
of the Roper resonance to $N\sigma$.
We are not aware of anything alike for mesons.

Before finishing our discussion,
let us again come back to the second possible solution
for the surface oscillations,
namely the much slower ones, with a value of $38\pm 4$ MeV.
Now, for nucleons we observed \cite{PRD27p1527} an average radial level
spacing about 12\% smaller than that for mesons,
and so an equally smaller oscillation frequency.
By AdS confinement, this also implies a larger hadron size.
So we might assume that surface oscillations for nucleons --- and
more generally baryons --- are of the order of $\omega =33\pm 4$ MeV.
The lowest surface excitation of a nucleon
would then have an excess energy of about
$\frac{1}{2}\omega =16.5$ MeV.
For isolated nucleon, the decay width of such excitation
could easily be too large to be observed.
Nowever, inside the nucleus, where the excitation may jump
from one nucleon to another, it might survive a bit longer.
Actually, excitations with energies
of this order of magnitude have been observed
in the distant past \cite{PR71p3}.
Could it be that surface oscillations of the gluon cloud
were noticed long before the quark model had even been considered?

In conclusion, constants of nature are extremely important
to help master its phenomena.
In previous work we found that quarkonia
are well described by a frequency of 190 MeV,
independent of flavors.
Here we seem to have discovered a second constant
of strong interactions for quarkonia, namely
an interference phenomenon with a constant periodicity
of about 76 MeV.
For baryons, which have a different color-charge configuration,
we suspect this value to be about 12\% smaller.
Furthermore,
we argue that
the observed interference patterns
in the amplitudes of annihilation processes
may be related to the surface oscillations of gluons.
From the intriguing fact that the observed periodicities
do not depend on the beam energy,
we find an additional indication for
flavor independence of the quarkonium oscillation frequency.

We are grateful for the precise measurements
and data analyses of the BABAR, CDF, and CMD-2 Collaborations,
which made the present analysis possible.
One of us (EvB) wishes to thank Dr. A.~A.~Osipov for useful discussions.
This work was supported in part by the {\it Funda\c{c}\~{a}o para a
Ci\^{e}ncia e a Tecnologia} \/of the {\it Minist\'{e}rio da Ci\^{e}ncia,
Tecnologia e Ensino Superior} \/of Portugal, under contract
CERN/\-FP/\-109307/\-2009.

\newcommand{\pubprt}[4]{#1 {\bf #2}, #3 (#4)}
\newcommand{\ertbid}[4]{[Erratum-ibid.~#1 {\bf #2}, #3 (#4)]}
\def\AP{Ann.\ Phys.}
\def\CPL{Chin.\ Phys.\ Lett.}
\def\JETPL{JETP Lett.}
\def\JPG{J.\ Phys.\ G}
\def\NCA{Nuovo Cim.\ A}
\def\PLB{Phys.\ Lett.\ B}
\def\PNL{Part.\ and Nucl.\ Lett.}
\def\PR{Phys.\ Rev.}
\def\PRC{Phys.\ Rev.\ C}
\def\PRD{Phys.\ Rev.\ D}
\def\PRL{Phys.\ Rev.\ Lett.}
\def\PREP{Phys.\ Rept.}
\def\ZPC{Z.\ Phys.\ C}

\end{document}